\documentclass[12pt]{article}
\usepackage{amsmath}
\usepackage{amssymb}
\usepackage{amsfonts}
\newcommand{\be}{\begin{equation}}
\newcommand{\ee}{\end{equation}}
\newcommand{\bea}{\begin{eqnarray}}
\newcommand{\eea}{\end{eqnarray}}
\newcommand{\nn}{\nonumber}
\newcommand{\p}{\phi}

\begin{document}
\title{Gauge symmetries of systems with a finite number of degrees of freedom }
\author{Farhang Loran\thanks{e-mail:
loran@cc.iut.ac.ir}\\ \\
  {\it Department of  Physics, Isfahan University of Technology,}\\
{\it Isfahan, 84156-83111, Iran}}
\date{}
  \maketitle
 \begin{abstract}
 For systems with a finite number of degrees of freedom, it is shown in \cite{non} that first class constraints  are Abelianizable if the Faddeev-Popov
 determinant is not vanishing for some choice of subsidiary constraints. Here, for irreducible first class constraint systems with SO(3) or SO(4) gauge
 symmetries, including a subset of coordinates in the fundamental
 representation of the gauge group, we explicitly determine the Abelianizable and
 non-Abelianizable classes of constraints.
 For the Abelianizable class, we explicitly solve the constraints to obtain the equivalent set of Abelian first class
 constraints. We show that for non-Abelianizable constraints there exist
  residual  gauge symmetries which results in  confinement-like phenomena.
 \end{abstract}

 \section{Introduction}
 Gauge theories can be understood as constraint systems with
 first class constraints which are the generators of gauge
 transformation \cite{Dirac}. In the Dirac method of quantization, physical states
 are, by definition, invariant under gauge transformation.
 In gauge fixing approaches \cite{Henbook,Govbook,GF} like the Faddeev-Popov method \cite{FP}, one
 eliminates the gauge freedom by introducing subsidiary constraints
 for which the Faddeev-Popov determinant  is not vanishing.
 These methods are equivalent to Dirac quantization as they are believed to
 generate an equivalent set of physical observables.

 Any given set of
 constraints $\{\phi_a\}$ can be replaced with a new set, say $\{\psi_a\}$, that is obtained by an
 invertible map from the original one. In this case, one says that $\{\phi_a\}$ and $\{\psi_a\}$ are {\em equivalent}. Usually, such a map is given
 as follows,
 \be
 \psi_a=\sum_{b=1}^A C_{ab}\phi_b,\hspace{1cm}a=1,\cdots, A'.
 \label{equ}
 \ee
 where $A$ and $A'$ are the cardinality of the sets $\{\phi_a\}$ and $\{\psi_a\}$ respectively and $C_{ab}$ are some functions of phase-space coordinates,
 which are not vanishing on the
 constraint surface. The set ${\phi_a}$ is {\em irreducible} if any
 equivalent set of constraints has the same cardinality, i.e.
 $A'=A$.

 In the case of gauge theories with a finite number of degrees of freedom, it is known that an irreducible set of first class class constraints is
 {\em Abelianizable} if
 there exists a set of subsidiary constraints such that the Faddeev-Popov determinant is not vanishing
 \cite{Henbook,non,abelable}. By an Abelianizable set of constraints
 $\phi_a$ one means a set of constraints that is equivalent to a new
 set of constraints $\{\psi_a\}$ with the Poisson algebra
 $\{\psi_a,\psi_b\}=0$.

 Thus, in the case of non-Abelianizable first class constraints,
 which are the generators of gauge transformation in non-Abelian gauge theories, the Faddeev-Popov determinant is vanishing for any choice of
 gauge fixing conditions \cite{non}.

 The proof is as follows: Consider a system with phase space
 coordinates $z_\mu$, $\mu=1,\cdots,2N$, and a set of  first class constraints
 $\p_a$, $a=1,\cdots,A\le N$ satisfying the algebra,
 \be
 \{\p_a,\p_b\}=f_{abc}\p_c,
  \ee
 where $\{\ ,\ \}$ stands for the Poisson bracket. Repeated indices are summed over.
 If $\p_a$'s are non-Abelianizable, one can prove that
 $\left(\frac{\partial \p_a}{\partial z_\mu}\right)$
 is not full rank  and consequently, as is stated above,
 the Faddeev-Popov determinant $\det(\{\p_a,\omega_b\})$ is vanishing for any
 choice of subsidiary constraints $\omega_a$, $a=1,\cdots,A$ \cite{non}. On the
 other hand if there exist a set of subsidiary constraints for which
 the Faddeev-Popov determinant is not vanishing, then one concludes
 that there exist a set of Abelian constraints equivalent to
 $\p_a$'s \cite{Henbook,non,abelable}. The proof given in
 \cite{Henbook}, is simple to follow: if for some set of subsidiary constraints the Faddeev-Popov determinant is not vanishing, then there exist
 at least one set of $A$ coordinates ${\tilde z}^a\in\{z^\mu\}$ for which
 $\det \left(\frac{\partial\p^a}{\partial {\tilde z}^b}\right)\neq
 0.$ Thus one can solve the constraints $\p^a=0$ for ${\tilde z}$ to obtain a set of new {\em equivalent} constraints $\psi^a={\tilde z}^a-f^a(z')=0$,
 in which by $z'$ one denotes the set of phase space coordinates complementary to $\tilde
 z$. It is now easy to show that $\psi^a$'s are Abelian constraints. Indeed the Poisson brackets of new constraints with each
 other as given as follows,
 \be
 \{\psi^a,\psi^b\}=\{{\tilde z}^a,{\tilde z}^b\}-\{{\tilde
 z}^a,f^b(z')\}-\{f^a(z'),{\tilde z}^b\}+\{f^a(z'),f^b(z')\},
 \label{j}
 \ee
 is independent of ${\tilde z}$'s since $\{{\tilde z}^a,{\tilde
 z}^b\}=0,\pm1$. On the other hand the right hand side of
 Eq.(\ref{j}) is vanishing on the constraint surface. Thus it
 vanishes identically \cite{Henbook}.

 Now consider a systems with a finite number of degrees of freedom
 with gauge group SO($N$) including $N$ coordinates, $q_a$,
 $a=1,\cdots, N$ in the fundamental representation. The first class
 constraints for such systems has the following general from,
 \be
 \phi_a=f_{abc}q^a p^b+L_a(q_i,p_i),
 \label{01}
 \ee
 where, $f_{abc}$ is the structure constant of ${\bf so}(N)$  algebra, $p_a$'s are momenta conjugate to $q_a$'s and $L_a$ are some
 functions of the other coordinates of systems and the corresponding
 momenta. Obviously, $L_a$'s are generators of gauge transformation in the
 subspace of phase space spanned by $q_i$'s and $p_i$'s, i.e.
 $\{L_a,L_b\}=f_{abc}L_c$. Consequently the space of gauge orbits
 factorizes as $O_a\otimes O_i$.

 {\bf Theorem} {\em${\bf so}(N)$  constraints are Abelianizable precisely if $L_a\neq0$ for some
 $a$.}

 {\bf Proof} Assume $L_1\neq 0$. The Faddeev-Popov determinant
 for the subsidiary constrains,
 \be
 \omega_a=\left\{\begin{array}{lll} q_a,&& a=1,\cdots,
 N-1,\\p_1,&&a=N,
 \end{array}\right.
 \label{02}
 \ee
  is not vanishing on the constraint surface,
 \be
 \det\left(\{\omega_a,\phi_b\}\right)=-(q_N)^{N-2} L_1\det f^{(N)},
 \label{03}
 \ee
 where $f^{(N)}_{ab}=f_{Nab}$.

 It is needless to say that the
 subsidiary constraints  $q_a=0$, $a=1,\cdots, N$, for which the Faddeev-Popov determinant is
 vanishing are not {\em suitable} gauge fixing conditions as they do not remove the
 gauge freedom. In fact, since the space of gauge orbits is factorized as $O_a\otimes O_i$ and the point $q_1=\cdots=q_N=0$ in $O_a$ is
 stationary under SO($N$) gauge
 transformation, in order to study gauge orbits and gauge transformations, one can
 concentrate on the space $O_a\backslash\{0\}\otimes O_i$. An open covering of the
 this space is given by open sets $U_a$ in which $q_a\neq 0$. The subsidiary constraints (\ref{02}) are gauge fixing
 conditions in the open set $U_N$.

 Since the constraints $\phi_a$ are Abelianizable, it is interesting to solve them explicitly and obtain the equivalent  set of Abelian constraints.
 For this, it suffices to find the Abelian constraints for  $U_N$ where $q_N\neq
 0$.  This appears to be a difficult task for general SO($N$) gauge group, though formal solutions of such equations are given in
 \cite{abelable}.

 In this paper, in sections 2 and 3, we calculate the explicit form of Abelian constraints
 for SO(3) and SO(4) cases respectively. In section 4, we study residual gauge symmetries in systems with non-Abelianizable first class constraints,
 and consider the discrete version of the Georgi-Glashow
 model \cite{Govpap} in which we obtain a simple confinement. Results are summarized in section
 5. In appendix A, we study Abelianization of
 constraints in a discrete version of the Higgs sector of the
 standard model.
 \section{Abelianization of SO(3) constraints}
 Consider a system with ${\bf so}(3)$ gauge algebra given by the following first class constraints,
 \bea
 \label{a1}
 \p_1&=&q_2p_3-q_3p_2+L_1,\\\
 \label{a2}
 \p_2&=&q_3p_1-q_1p_3+L_2,\\
 \label{a3}
 \p_3&=&q_1p_2-q_2p_1+L_3.
 \eea
 where $L_a$ ($a=1,2,3$) is not a function of  $q_a$'s and $p_a$'s.

 If $q_3\neq0$, the constraints given in Eqs.(\ref{a1}-\ref{a3})
 are equivalent to the following constraints,
 \bea
 \label{f1}
 \psi_1&=&{\p_1\over q_3}=p_2-\frac{q_2}{q_3}p_3-\frac{L_1}{q_3},\\
 \label{f2}
 \psi_2&=&{\p_2\over q_3}=p_1-\frac{q_1}{q_3}p_3+\frac{L_2}{q_3},\\
 \label{f3}
 \psi_3&=&\sum_{a=1}^3q_a\phi_a=q_1L_1+q_2L_2+q_3L_3.
 \eea
  It is easy to verify that $\{\psi_{1(2)},\psi_3\}=0$, and
 \be
 \{\psi_1,\psi_2\}=-\frac{\psi_3}{(q_3)^3}.
 \label{g1}
 \ee
 To make the right hand side of Eq.(\ref{g1}) vanishing, $\psi_2$ can be
 redefined as follows,
 \be
 \psi_2\to\psi_2^{\rm new}=\psi_2^{\rm old}-\frac{q_2}{q_3}\frac{1}{(q_2)^2+(q_3)^2}\psi_3.
 \label{k}
 \ee
  it is easy to show that $\{\psi_{1(3)},\psi_2^{\rm new}\}=0$.


  \section{Abelianization of SO(4) constraints}
  Consider first class constraints in ${\bf so}(4)$ gauge algebra,
  \be
  \phi_a=f_{abc}q_ap_c+L_a
  \label{t1}
  \ee
  where the non-vanishing structure coefficients are
  \be
  f_{321}=f_{156}=f_{246}=f_{345}=1,
  \label{t2}
  \ee
  we assume that $q_1\neq 0$ and solve the constraints to obtain six
  equivalent Abelian constraints. First we replace $\phi_1$ with
  $\psi_1$ defined as follows,
  \be
  \psi_1=\sum_{a=1}^6q_a\phi_a=\sum_{a=1}^6q_aL_a.
  \label{t3}
  \ee
  One can easily verify that the Poisson bracket of the other five
  constraints with $\psi_1$ is vanishing. Since $q_1\neq 0$, one can
 solve constraints for $p_a$ ($a=2,3,5,6$) in terms of $p_1$ and
 $p_4$. It should be noted that by {\em solving constraints} one obtains a new set of constraints that are equivalent to the original ones in the
 sense of Eq.(\ref{equ}). Solving $\phi_2=0$ one obtains,
 \be
 p_3={q_3\over q_1}p_1-{q_4\over q_1}p_6+{q_6\over
 q_1}p_4-{L_2\over q_1},
 \label{t4}
 \ee
 and $\phi_5=0$ gives,
 \be
 p_6={q_6\over q_1}p_1-{q_4\over q_1}p_3+{q_3\over
 q_1}p_4+{L_5\over q_1}.
 \label{t5}
 \ee
 Eqs.(\ref{t4}) and (\ref{t5}) can be solved to obtain,
 \bea
 \label{t6}
 \left(1-\left(q_4\over q_1\right)^2\right)p_3=\left({q_3\over
 q_1}-{q_4q_6\over q_1^2}\right)p_1+\left({q_6\over
 q_1}-{q_4q_3\over q_1^2}\right)p_4-\left({L_2\over q_1}+{q_4L_5\over
 q_1^2}\right),\\
 \label{t7}
 \left(1-\left(q_4\over q_1\right)^2\right)p_6=\left({q_6\over
 q_1}-{q_4q_3\over q_1^2}\right)p_1+\left({q_3\over
 q_1}-{q_4q_6\over q_1^2}\right)p_4+\left({L_5\over q_1}+{q_4L_2\over
 q_1^2}\right).
 \eea
 For a generic point on the phase space, $q_1\neq q_4$ and
 consequently constraints $\phi_2=0=\phi_5$ are equivalent to the
 following new constraints,
  \bea
 \label{t8}
 \psi_3=p_3-{q_3q_1-q_4q_6\over{q_1^2-q_4^2}}p_1-{q_6q_1-q_4q_3\over{q_1^2-q_4^2}}p_4+{q_1L_2+q_4L_5\over{q_1^2-q_4^2}},\\
 \label{t9}
 \psi_6=p_6-{q_6q_1-q_4q_3\over{q_1^2-q_4^2}}p_1-{q_3q_1-q_4q_6\over{q_1^2-q_4^2}}p_4-{q_1L_5+q_4L_2\over{q_1^2-q_4^2}}.
 \eea
 Similarly one can show that constraints $\phi_3=0=\phi_6$ are
 equivalent to the following constraints,
 \bea
 \label{t10}
 \psi_2=p_2-{q_2q_1+q_4q_5\over{q_1^2-q_4^2}}p_1+{q_5q_1+q_4q_2\over{q_1^2-q_4^2}}p_4-{q_1L_3-q_4L_6\over{q_1^2-q_4^2}},\\
 \label{t11}
 \psi_5=p_5-{q_5q_1+q_4q_2\over{q_1^2-q_4^2}}p_1+{q_2q_1+q_4q_5\over{q_1^2-q_4^2}}p_4+{q_1L_6-q_4L_3\over{q_1^2-q_4^2}}.
 \eea
  We define  $\psi_4$ by solving $\phi_4$ in terms of $p_1$ and $p_4$
  using the above constraints,
  \be
  \psi_4=q_1L_4+q_4L_1-q_2L_5-q_5L_2+q_6L_3+q_3L_6.
  \label{t12}
  \ee
  It is  straightforward to show that
  $\{\psi_a,\psi_1\}=0=\{\psi_a,\psi_4\}=0$, ($a=1,\cdots,6$).
  Furthermore, one can show that,
  \be
  \begin{array}{l}
   \{\psi_2,\psi_5\}=\{\psi_3,\psi_6\}=0,\\
   \{\psi_3,\psi_2\}=\{\psi_5,\psi_6\}=R_1\psi_1-R_2\psi_4,\\
   \{\psi_2,\psi_6\}=\{\psi_3,\psi_5\}=R_2\psi_1-R_1\psi_4,
   \end{array}
   \label{t14}
   \ee
   where
   \be
   R_1={{q_1^3+3q_1q_4^2}\over{\left(q_1^2-q_4^2\right)}^3},\hspace{1cm}R_2={{q_4^3+3q_4q_1^2}\over{\left(q_1^2-q_4^2\right)}^3}.
   \label{t15}
   \ee
 It is straightforward to show that by replacing $\psi_3$ and
 $\psi_6$ with the following equivalent constraints,
 \bea
 \psi_3\to\psi^{\rm new}_3=\psi^{\rm old}_3+S_1\psi_1-S_2\psi_4,\nn\\
 \psi_6\to\psi^{\rm new}_6=\psi^{\rm old}_6-S_2\psi_1+S_1\psi_4,
 \label{t16}
 \eea
 \bea
 S_1=\frac{1}{2}\left[\frac{(q_2+q_5)/(q_1-q_4)}{(q_1-q_4)^2+(q_2+q_5)^2}+
 \frac{(q_2-q_5)/(q_1+q_4)}{(q_1+q_4)^2+(q_2-q_5)^2}\right],\nn\\
 S_2=\frac{1}{2}\left[\frac{(q_2+q_5)/(q_1-q_4)}{(q_1-q_4)^2+(q_2+q_5)^2}-
 \frac{(q_2-q_5)/(q_1+q_4)}{(q_1+q_4)^2+(q_2-q_5)^2}\right].
 \label{t17}
 \eea
 one can obtain a new set of constraints which are Abelian.
 \subsection{Equivalence of Constraints}
 Two sets of first class constraints are equivalent precisely if the corresponding constraint surfaces
 and gauge transformations are equivalent. In the case studied here, the constraints surfaces of the SO(4) and the
 Abelian constraints are equivalent by construction, since the Abelian constraints are found
 by solving the SO(4) constraints. A possible flaw might be at the intersection of the constraint surface of
 $\phi_a$ given in Eq.(\ref{t1}) and the $q_1^2-q_4^2=0$ surface. For example, Eqs.(\ref{t4}) and (\ref{t5}) imply that
 at $q_1^2-q_4^2=0$,
 \bea
 \label{t18}
 p_3+p_6&=&{q_3\over q_1}p_1+{q_6\over q_1}p_4-{L_2\over q_1},\\
 \label{t19}
  p_6+p_3&=&{q_6\over q_1}p_1+{q_3\over q_1}p_4+{L_5\over q_1}.
 \eea
 Thus, it is necessary to see  whether the constraints $\psi_3$ and $\psi_6$ given in
 Eqs.(\ref{t8}) and (\ref{t9}) give Eqs.(\ref{t18}) and (\ref{t19}) at
 $q_1^2-q_4^2=0$. To deal with this problem, let's assume that for
 example,
 $q_1=q_4+\epsilon$ where $\epsilon$ is an infinitesimal parameter. By this assumption, Eqs.(\ref{t8}) and
 (\ref{t9}) give the following equations,
 \bea
 \label{tt8}
 2\epsilon q_1p_3&=&\left(q_3q_1-q_4q_6\right)p_1+\left(q_6q_1-q_4q_3\right)p_4-\left(q_1L_2+q_4L_5\right),\\
 \label{tt9}
 2\epsilon q_1p_6&=&\left(q_6q_1-q_4q_3\right)p_1+\left(q_3q_1-q_4q_6\right)p_4+\left(q_1L_5+q_4L_2\right).
 \eea
 Eq.(\ref{tt8}) for $\epsilon\to0$ gives,
 \be
 \left(q_3-q_6\right)p_1+\left(q_6-q_3\right)p_4-\left(L_2+L_5\right)=0
 \label{tt10}
 \ee
 Furthermore, by adding the left and right sides of Eqs.(\ref{tt8}) and (\ref{tt9}) one obtains,
 \be
 2\epsilon q_1\left(p_3+p_6\right)=\epsilon\left[\left(q_3+q_6\right)p_1+\left(q_6+q_3\right)p_4-\left(L_2-L_5\right)
 \right].
 \label{tt11}
 \ee
 It is clear that Eqs.(\ref{tt11}) and (\ref{tt10}) give
 Eq.(\ref{t18}). Eq.(\ref{t19}) can be obtained in the same way and
 furthermore  all these consistency checks can be done for the case
 $q_1=-q_4+\epsilon$ and $\epsilon\to 0$.

 The above method of calculations motivates us to introduce an
 infinitesimal parameter in the denominators as follows,
 \be
 {1\over q_1^2-q_4^2}\to {1\over q_1^2-q_4^2+i\epsilon},
 \label{rule}
 \ee
 and consider a rule for calculations: setting $\epsilon$ to zero is the final step in all
 calculation. The same rule resolves the ambiguity in the definition
 of functions $S_1$ and $S_2$ in Eq.(\ref{t17}).

 To verify the equivalence of Abelian gauge transformations and the SO(4) gauge transformations,
 let $\theta_a$ and $\eta_a$ ($a=1,\cdots,6$) be the gauge
 parameters corresponding to the SO(4) and the Abelian gauge transformations
 respectively.
 $\theta_a$ and $\eta_a$ are in general functions of
 phase space coordinates. The gauge transformation of a function of
 phase space coordinates like $F$ is given as follows:
 \bea
 \label{e1}
 \delta^{\rm A}F&=&\sum_{a=1}^6\theta_a\{F,\phi_a\},\\
 \label{e2}
 \delta^{\rm nA}F&=&\sum_{a=1}^6\eta_a\{F,\psi_a\}.
 \eea
 The parameters $\eta_a=\delta^{\rm A}x_a$ ($a=2,3,5,6$) can be determined in terms
 of the parameters $\theta_a$ by the condition $\delta^{\rm
 A}x_a=\delta^{\rm nA}x_a$ for $a=2,3,5,6$. A nontrivial observation is that
 these conditions give also $\delta^{\rm A}x_a=\delta^{\rm nA}x_a$ for
 $a=1,4$. This means that the Abelian and SO(4) gauge transformations of coordinates $x_a$ are precisely equivalent.
 $\eta_1$ and $\eta_4$ can be determined after a lengthy but quite straightforward calculation by examining the
 gauge transformation of, say, $p_3$.

 In general, $\delta^{\rm A}F\approx\delta^{\rm nA}F$
 where the symbol of {\em weak equality}  $\approx$ means equality on the constraint surface \cite{Dirac}. This
 is good because for constraint systems, physical quantities are defined on the constraint
 surface. The non-trivial result above was the precise equivalence of the SO(4) and the Abelian gauge transformations
 for $x_a$'s.
 \section{Residual $U(1)$ gauge symmetry}
 Now we deal with the class of systems with non-Abelianizable
 constraints. These are constraint systems for which $L_a=0$ in
 Eq.(\ref{01}).

 {\bf Lemma} {\em For constraints,
 \be
 \phi_a=f_{abc}q^a p^b.
 \ee
 there exist one non-gaugeable residual $U(1)$ gauge symmetry generated
 by}
 \be
 \psi_1=\sum_aq_a\phi_a.
 \ee
 As we have seen, for $SO(4)$ constraints there exist another $U(1)$
 residual gauge symmetry generated by $\psi_4$ given in
 Eq.(\ref{t12}).

 The existence of residual gauge symmetries in systems with
 non-Abelianizable first class constraints is a consequence of the main
 theorem in \cite{non} as can be seen as follows.

 {\bf Corollary} {\em In a gauge theory with  non-Abelianizable gauge symmetry,
  any classical configuration $z_\mu=z_\mu^v$ is invariant under a non-trivial subgroup of the
 non-Abelian gauge group.}

 To see this, let's define the generators of
 gauge transformation by
 \be
 \delta_a \eta(z_\mu)=\{\eta,\p_a\}
 \ee
  in which $\eta(z_\mu)$ is any function of the phase space coordinates.
 Define $\lambda_i^a$, $i=1,\cdots,I$ to be the
 $i$-th null vector of $\left(\frac{\partial \p_a}{\partial
 z_\mu}\right)_{z_\mu^v}$, and define $\delta_i=\lambda_i^a\delta_a$.  Now it is easy to verify that
 $\left(\delta_i\eta\right)_{z^v}=0$ for any function $\eta(z)$. We
 rearrange the $A$ generators of gauge transformation to $I$
 $\delta_i$'s and the complementary set $\delta_\alpha$ where the
 index $\alpha$ runs over $1,\cdots,A-I$. One can consider $A-I$
 subsidiary constraints which gauge the gauge freedom corresponding
 to $\delta_\alpha$'s. But since $\delta_i\eta|_{z^v}=0$ for any function $\eta$, there is
 no way to gauge the gauge symmetry generated by $\delta_i$'s.
 Recall that a gauge fixing condition is a function $\omega$ which
 is  not invariant under the gauge transformations.
 Consequently, any classical configuration of the system is
 invariant under the gauge transformation generated by
 $\delta_i$'s.
 \subsection{Quantization} To deal with $\delta_i$'s,
 the only consistent method of quantization is to use the Dirac definition of
 physical states. Thus after imposing the $A-I$ possible gauge fixing conditions to gauge $\delta_\alpha$'s,
 one defines/assumes the
 physical state to be invariant under $H_i$'s which are the quantum
 operators corresponding
 to the classical generators $\delta_i$'s. It is a natural assumption since $\delta_i$'s are by definition
 the symmetries of the classical configurations. This implies that the only physical observables are those combinations
 of field operators that are invariant under $H_i$'s. This is again in agreement with the classical result
 $\delta_i\eta|_{z^v}=0$. This phenomena can be interpreted as
 confinement.

 We state without proof the following conjecture.

{\bf Conjecture:} {\em $H_i$'s are the generators of the Cartan
 subalgebra of the gauge group $G$ generated by $\p_a$'s and consequently $I$ is equal to the rank of the gauge
 group.}

 If this conjecture is valid then one verifies that the number of
gauge symmetries that can be fixed by
 gauge fixing conditions equals the number of non zero roots of the gauge
 group.
 \subsection{Example} Here we give an illustrative simple examples which shed some light on different aspects of the general arguments and statements
 given above.

 This is the discrete version of the Georgi-Glashow
 model in which we obtain a simple confinement. The model is
 given by the Lagrangian
 $L=\frac{1}{2}{\dot{\vec q}}^2-V({\vec q})$, which is invariant under the action of $SO(3)$
 \cite{Govpap}:
 \be
 q\to g q,\hspace{1cm}g=e^{i \theta \hat{n}.\phi},\ \phi\in so(3).
 \ee
 For example, $V({\vec q})=({\vec q}^2-a^2)^2$.
 It is easy to verify that any classical vacuum ${\vec q}^v$ ($\left|{\vec q}^v\right|=a$),
 is invariant under the $U(1)$ subgroup of $SO(3)$ generated by ${\hat
 q}^v.\phi$ as is expected. In fact $\delta_{\epsilon,{\hat n}} q_i^v=\epsilon ({\hat
 n}\times {\vec q}^v)_i$ which is vanishing if ${\hat n}=\hat{q}^v$.

  To make connection between this seemingly trivial result and the general arguments
 given above, let's consider a gauge field ${\rm \bf A}$ in the adjoint representation of
 $so(3)$ and the Lagrangian $L=\frac{1}{2}({D_t{\vec q}})^2-V({\vec
 q})$, where $D_t {\vec q}={\dot{\vec q}}+{\rm \bf A}{\vec q}$.  The
 corresponding Hamiltonian is $H=H_0+A_iL_i$ where
 $A_i=\epsilon_{ijk}{\rm \bf A}_{ij}$, $\phi_i=\epsilon_{ijk}q_jp_k$ and $H_0=\frac{1}{2}{\vec p}^2+V({\vec
 q})$.
 ${\vec p}={\dot{\vec q}}+{\rm \bf A}{\vec q}$ is the momentum
 conjugate to $\vec q$. The momenta
 conjugate to the gauge field $A$ are vanishing. These are the primary first class constraints. The corresponding
equations of motion
 result in the secondary
 first class constraints  $\phi_i=0$. The secondary constraints here are the
 generators of the gauge group $SO(3)$ as they satisfy the algebra,
 $\{\phi_i,\phi_j\}=\epsilon_{ijk}\phi_k$. One can easily show that $\hat{q}^v$ is the unique
 null vector of
 \be
 \left(\frac{\partial \phi_i}{\partial
 z_\mu}\right)_{z^v_\mu}=\left(\begin{array}{cccccc}
 0&0&0&0&-q^v_3&q^v_2\\
 0&0&0&q^v_3&0&-q^v_1\\
 0&0&0&-q^v_2&q^v_1&0
 \end{array}\right),
 \label{so3}
\ee
 where $z_\mu=({\vec q},{\vec p})$ are the phase space coordinates and
 $z_\mu^v=({\vec q}^v,{\vec 0})$ is a classical vacuum state.
 Therefore the vacuum ${\vec q}^v$ is invariant under $U(1)\subset
 SO(3)$ generated by ${\hat q}^v.L$.

 What is the confinement in this
 example? We argued that in general the maximum number of gauge degrees of freedom that can be gauged equals
${\rm rank}\left(\frac{\partial \p_a}{\partial z_\mu}\right)=A-I$
 in which $A$ is the number of the gauge generators.
 From Eq.(\ref{so3}) one verifies that $I=1$ and consequently one can impose at most two
 gauge fixing conditions. Let's assume that these two subsidiary constraints are $q_3=0=p_3$.
 Namely we are assuming that the trajectory of the particle is in
 the 1-2 plane.
 Since the total angular momentum is vanishing the trajectory is a straight line which can be assumed
 to pass through the origin without loss of
 generality.
 The $U(1)$ symmetry here is the symmetry under arbitrary rotation
 of this line around the third axis. Let's define new coordinates
 $z=q_1+iq_2$ and ${\bar z}=q_1+iq_2$, which under the $U(1)$
 transformation change a phase, $z\to e^{-i\wp}z$ and ${\bar
 z}\to e^{i\wp}{\bar z}$. Assuming that the vacuum state $\left|0\right>$ corresponding to the classical vacua ${\vec q}^v=(0,0,q^v)$ is
 invariant under symmetries of the classical vacua, one verifies that e.g.  $\left<z\right>=0$ while $\left<z{\bar
 z}\right>$ can be in general non-vanishing. Considering $z$ as a {\em quark}, this observation can be interpreted as confinement.


 \section{Summary}
 For first class constraint systems with first class constraints
 \be
 \phi_a=f_{abc}q^a p^b+L_a(q_i,p_i),
 \label{con}
 \ee
 satisfying constraint algebra,
 \be
 \{\phi_a,\phi_b\}=f_{abc}\phi_c,
 \ee
 in which $f_{abc}$ is the structure coefficients of SO(3) or SO(4) Lie algebras, and at least one $L_a\neq 0$, we obtained the equivalent set
 of Abelian first class constraints.

 For {\bf so}(3) gauge algebra, with structure coefficient $f_{abc}=\epsilon_{abc}$ the Abelian constraints for the $q_3\neq 0$ subset of phase
 space are given as follows,
 \bea
 \psi_1&=&p_2-\frac{q_2}{q_3}p_3-\frac{L_1}{q_3}\nn\\
 \psi_2&=&p_1-\frac{q_1}{q_3}p_3+\frac{L_2}{q_3}-\frac{q_2}{q_3}\frac{1}{(q_2)^2+(q_3)^2}\psi_3\nn\\
 \psi_3&=&q_1L_1+q_2L_2+q_3L_3.
 \label{con1}\eea
 Appropriate transition functions will give the corresponding
 Abelian constraints in the $q_1\neq 0$ and $q_2\neq 0$ subsets of
 the phase space. We have excluded the point $q_1=q_2=q_3=0$, which
 is stationary under gauge transformations.

 For {\bf so}(4) gauge algebra, with  non-vanishing structure coefficients
 \be
 f_{321}=f_{156}=f_{246}=f_{345}=1,
 \ee
 the Abelian constraints in the $q_1\neq 0$ subset of phase space are,
 \be
 \begin{array}{l}
 \psi_1=\sum_{a=1}^6q_aL_a,\\\\
 \psi_2=p_2-{q_2q_1+q_4q_5\over{q_1^2-q_4^2+i\epsilon}}p_1+{q_5q_1+q_4q_2\over{q_1^2-q_4^2+i\epsilon}}p_4-{q_1L_3-q_4L_6\over{q_1^2-q_4^2+i\epsilon}},
 \\\\
 \psi_3=p_3-{q_3q_1-q_4q_6\over{q_1^2-q_4^2+i\epsilon}}p_1-{q_6q_1-q_4q_3\over{q_1^2-q_4^2+i\epsilon}}p_4+{q_1L_2+q_4L_5\over{q_1^2-q_4^2+i\epsilon}}
 +S_1 \psi_1-S_2\psi_4,\\\\
 \psi_4=q_1L_4+q_4L_1-q_2L_5-q_5L_2+q_6L_3+q_3L_6,\\\\
 \psi_5=p_5-{q_5q_1+q_4q_2\over{q_1^2-q_4^2}}p_1+{q_2q_1+q_4q_5\over{q_1^2-q_4^2+i\epsilon}}p_4+{q_1L_6-q_4L_3\over{q_1^2-q_4^2+i\epsilon}},\\\\
 \psi_6=p_6-{q_6q_1-q_4q_3\over{q_1^2-q_4^2+i\epsilon}}p_1-{q_3q_1-q_4q_6\over{q_1^2-q_4^2+i\epsilon}}p_4-{q_1L_5+q_4L_2\over{q_1^2-q_4^2+i\epsilon}}
 - S_2\psi_1+ S_1\psi_4,
 \end{array}
 \label{con2}
 \ee
 where $S_1$ and $S_2$ are defined in Eq.(\ref{t17}) and $\epsilon$ is a parameter which one sets to zero at the end of
 calculations. This parameter is introduced to resolve the apparent
 singularity at  $q_1^2-q_4^2=0$.

 For the non-Abelianizable constraints, which is the case with $L_a=0$ in Eq.(\ref{con}), there are exist
  residual  gauge symmetries which results in confinement-like phenomena.

 First class constraint systems with SO($N$) gauge symmetry generated by
 first class class constraints (\ref{con}) are interesting specially
 as toy models to study Gribov copies in non-Abelian gauge theories.
 Results given in Eqs.(\ref{con1}) and (\ref{con2}) can be used to study this problem from a new point of view.
 \appendix
 \section{The Higgs sector of the standard model}
 In this appendix, we study a special SO(4) invariant constraint
 system in which the {\bf so}(4) Lie algebra is represented by first class constraints
 constraints in a different way in comparison to section 3.

 We consider a system with 4 degrees
of freedom  $q_\alpha$, plus a gauge field. The index $\alpha$ runs
over 0,1,2,3. The Lagrangian is the following $L= 1/2
(\dot{q}_\alpha - A_i \eta^i_{\alpha \beta} q_\beta)^2 -V(q)$ where
Latin indices $i$ runs over 1,2,3. Repeated indices are summed over.
The potential can be taken to be $V(q)= \lambda (q_\alpha^2 -1)^2$.
The symbols $\eta^i_{\alpha \beta}$ are 't Hooft symbols,
 \be
 \eta^i_{\alpha\beta}=\epsilon_{0i\alpha\beta}-\delta_{i\alpha}\delta_{0\beta}+\delta_{0\alpha}\delta_{i\beta},
 \label{tooft}
 \ee
 satisfying the commutation relation,
 $\eta^{[i}_{\alpha_\rho}\eta^{j]}_{\beta_\rho}=2\epsilon_{ijk}\eta^k_{\alpha\beta}.$ This
 looks just like a discrete version of the Higgs sector of the
 standard model.

 The conjugate momentum to the gauge field $A_i$ vanishes. These are
 the primary constraints. The secondary first class constraints are
 obtained by differentiation with respect to $A_i$. They are $\phi_i= -
 p_\alpha \eta^i_{\alpha \beta} q_\beta$ where $p_\alpha =
 (\dot{q}_\alpha - A_i \eta^i_{\alpha \beta} q_\beta)$ is the
 conjugate momentum to $q_\alpha$. It is easy to see, using the
 expression of the 't Hooft symbols that $\{ \phi_i, \phi_j\} = 2
 \epsilon_{ i j k} \phi_k$. Thus $\phi_i$ are non-Abelian constraints
 generating a SU(2) subgroup of SO(4). Now one can introduce the following
 subsidiary constraints, which are equivalent to unitary gauge
 $q_i=0$. The Poisson brackets of these constraints with the $\phi_i$
 are $\{q_i, \phi_j\} = q_0 \delta_{ i j}$ which is non-vanishing for
 $q^0\neq 0$.

 We show that $\phi_i$'s are Abelianizable if $q_0\neq 0$. Thus we are
realizing two different sectors in the theory. In one sector $q_0$
and consequently $p_0$ are both vanishing as we will show in a
moment. Thus the SO(4) model reduces to  the SO(3) model studied in
section 2.  In the sector $q_0\neq 0$, we show that the secondary
constraints $\phi_i=0$ are equivalent to three Abelian constraints.

The proof is as follows. Using Eq.(\ref{tooft}), one can show that,
 \be
 \vec \phi=\vec q\times\vec p+q_o\vec p-p_0\vec q,
 \ee
 in which $\vec \phi=(\phi_1,\phi_2,\phi_3)$. The constraint $\vec \phi=0$ implies that
 \bea
 \vec q.\vec
 \phi=q_0\vec q.\vec p-p_0{\vec q}^2=0,\nn\\
 \vec p.\vec \phi=p_0\vec q.\vec p-q_0{\vec p}^2=0.
 \eea
 The cases with vanishing $\vec q^2$ or $\vec p^2$ are rather trivial.
 The most nontrivial cases are given either by $q_0=p_0=0$ or by $\vec \psi=q_0\vec p-p_0\vec
 q=0$ and $q_0\neq 0\neq p_0$. In the first case, one obtains the
 SO(3) model and $\det\left(\{q_i,\phi_j\}\right)=0$ whatever the gauge fixing conditions are. In the second case, the
 constraints $\phi_i=0$ are Abelianizable as they are equivalent to Abelian constraints $\psi_i=0$. $\phi_i$'s and $\psi_i$'s
 are equivalent as they define the same constraint surface in the phase space. But by ``equivalence" in \cite{non} one means also
 equivalence in the gauge transformation generated by two sets of first class constraints which we have not verified
 yet. The gauge transformation generated by $\phi_i$'s is given by
 $\delta \vec q=\{\vec q,\vec n.\vec \phi\}=\vec n\times \vec q+q_0\vec n$, where
 $\vec n$ is the parameter of gauge transformation. Since $q_0\neq
 0$ one can easily verify that $\delta\vec q=\{\vec q, \vec n'.\vec
 \psi\}$ in which $\vec n'=\vec n+q_0^{-1}\vec n\times \vec q$.

\end{document}